\documentclass[a4paper,11pt]{article}
\usepackage[a4paper, margin=25mm]{geometry}

\usepackage{authblk}
\usepackage{cite}
\usepackage[cmex10]{amsmath}
\usepackage{amssymb,amsfonts}
\usepackage{graphicx}
\usepackage{textcomp}
\usepackage{xcolor}
\usepackage{wrapfig}
\def\BibTeX{{\rm B\kern-.05em{\sc i\kern-.025em b}\kern-.08em
    T\kern-.1667em\lower.7ex\hbox{E}\kern-.125emX}}
    
\usepackage[utf8]{inputenc}

\usepackage{pgfplots}
\pgfplotsset{compat=1.15}

\usepackage{tikz}
\usetikzlibrary{calc,trees,positioning,matrix}
\usepackage{adjustbox}
\usepackage{geometry}
\usepackage{changepage}

\usepackage[caption=false,font=footnotesize]{subfig}

\usepackage{float}

\usepackage{booktabs}
\usepackage{pgffor}
\usepackage{hyperref}

\begin{document}

\title{Euler Meets GPU:\\Practical Graph Algorithms with Theoretical Guarantees\thanks{This work was partially supported by the Polish Ministry of Science and Higher Education grant DI2012~018942 and National Science Center of Poland grant 2017/27/N/ST6/01334.}}

\author{Adam Polak}
\author{Adrian Siwiec}
\author{Michał Stobierski}
\affil{Jagiellonian University}
\date{}
\maketitle

\begin{abstract}
The Euler tour technique is a classical tool for designing parallel graph algorithms, originally proposed for the PRAM model. We ask whether it can be adapted to run efficiently on GPU. We focus on two established applications of the technique: (1) the problem of finding lowest common ancestors (LCA) of pairs of nodes in trees, and (2) the problem of finding bridges in undirected graphs. In our experiments, we compare theoretically optimal algorithms using the Euler tour technique against simpler heuristics supposed to perform particularly well on typical instances. We show that the Euler tour-based algorithms not only fulfill their theoretical promises and outperform practical heuristics on hard instances, but also perform on par with them on easy instances.
\end{abstract}

\section{Introduction}

Over the last two decades graphical processing units (GPU) proved very successful for general purpose computation. 
Their architecture makes them particularly well suited for highly parallel jobs such as matrix multiplication and other linear algebraic operations. Hence, the recent neural networks boom further increased the GPU popularity. While matrix multiplication is an embarrassingly parallel problem -- and solving it efficiently with GPU requires more of an engineering than algorithmic insight -- for other less regular problems we face an obstacle: the GPU architecture is complex and has no good theoretical model.\footnote{Such as the word-RAM model, which proved to be a good model for modern CPU, in the sense that it often facilitates developing practical algorithms.} In order to design an effective GPU algorithm, performing well in practice, one needs to take care of certain low-level engineering aspects which cannot be abstracted away, e.g., different synchronization at different levels of parallelization (i.e.~\emph{threads} and \emph{blocks}), or high latency global memory, which requires overlapping reads to achieve a good throughput.

Probably the closest model is the \emph{parallel random-access machine} (PRAM) model, developed in the 1980s. Certain features make it though extremely unrealistic, e.g.~zero-cost synchronization, or zero-latency memory. It thus serves more as a source of inspiration for GPU algorithms rather than an actual model. Some PRAM algorithms were successfully adapted to GPU (e.g.~prefix sum~\cite{Scan07}), but some are believed to be too complex and to have too large constant overheads to become practical on GPU (see e.g.~\cite{EV12,WK17} on the biconnectivity algorithm of~\cite{TV85}).

A common approach for GPU algorithm design is to use simple low-overhead algorithms that exploit parallelism assumed to be present in typical real-world instances (but not guaranteed by a combinatorial structure of the problem). This is exemplified by arguably the most popular GPU graph primitive, \emph{breadth-first search} (BFS). GPU implementations of BFS (e.g.~\cite{MGG12, Gun17}) are extremely efficient on graphs with small diameter -- which is the case for many real-world graphs -- but become prohibitively slow on, say, paths.

The Euler tour technique is a classical tool for designing parallel graph algorithms, originally proposed for the PRAM model~\cite{TV85}. We ask whether it can be adapted to run efficiently on GPU. We focus on two established applications of the technique: (1) the problem of finding \emph{lowest common ancestors} (LCA) of pairs of nodes in trees, and (2) the problem of finding \emph{bridges} in undirected graphs. In our experiments, we compare theoretically optimal algorithms using the Euler tour technique~\cite{SV88,TV85} against simpler heuristics~\cite{MT12,WK17}, which are state-of-the-art GPU algorithms, supposed to perform particularly well on typical instances. We show that the Euler tour-based algorithms not only fulfill their theoretical promises and outperform practical heuristics on hard instances, but also perform on par with them on easy instances.

\subsection{Related Work}

\subsubsection*{Euler Tour} Tarjan and Vishkin~\cite{TV85} first introduced the Euler tour technique, and applied it to design a parallel biconnectivity algorithm.
Later the technique was used in many PRAM algorithms from different areas, e.g.~algorithms on trees~\cite{CV88, GIR88}, planar graphs~\cite{Gazit91, RR94}, polygons~\cite{Goodrich96}, texts~\cite{CR90, CAL10}, and context-free grammars~\cite{Rytter86}. It also found applications beyond PRAM, e.g.~in large-scale distributed graph processing~\cite{Pregel14,AMPC19}, and dynamic problems~\cite{Miltersen94,Henzinger95,Tarjan97}.

\subsubsection*{Graph Algorithms on GPU}
First attempts to speed up graph algorithms using GPU included transitive closure~\cite{Larsen01} and APSP~\cite{Mici04}. Harish et al.~\cite{Harish07,Harish09} took a systematic approach and developed GPU implementations of BFS, ST-connectivity, shortest paths, minimum spanning tree, and max-flow algorithms. Other examples include triangle counting~\cite{Wang16}, connected components~\cite{Soman10,JB18}, and strongly connected components~\cite{Barnat11,Li17}. Most of those works report speedups in the range of 10-100x over single-core CPU baselines. They also make it clear that developing from scratch an efficient GPU algorithm, which can outperform a baseline, often requires a significant amount of nontrivial fine-tuning. For that reason, there is a growing interest in graph processing libraries for GPU, delivering highly-optimized building blocks allowing an easy construction of efficient algorithms, e.g.~Medusa~\cite{Medusa14}, Gunrock~\cite{Gun17}, Groute~\cite{Groute17}.

\subsubsection*{LCA in Trees}
Aho, Hopcroft and Ullman~\cite{AH73} were the first to study the LCA problem. On top of being a natural combinatorial problem, LCA was also extensively studied as a subproblem emerging in other graph problems, e.g., dominator tree in a directed flow-graph~\cite{AH73}, maximum weighted matching~\cite{G90}, network routing~\cite{TZ01}, phylogenetic distance computation~\cite{MT12}. Harel and Tarjan~\cite{HT84} first gave an optimal LCA algorithm, with linear time preprocessing and constant time queries. It was subsequently simplified by Schieber and Vishkin~\cite{SV88}, and Bender and Farach-Colton~\cite{BFC00}. The approach of Schieber and Vishkin~\cite{SV88} is not only simple but also naturally parallel, yielding an optimal, $O(\log n)$-time and $O(n)$-work, PRAM algorithm.

Martins et al.~\cite{MT12} implemented an LCA algorithm on GPU. They opted for a na\"ive approach, working well on shallow trees, but with a linear, instead of a constant, worst-case query time. To the best of our knowledge, the only GPU alternative to their algorithm is the work of Soman et al.~\cite{Soman10b}, which focuses on discrete range searching primitives, and mentions LCA only briefly as one of applications. They assume, however, that a required preprocessing is already done and the necessary data structures are given in input, and they focus only on an algorithm for answering queries. Therefore, we cannot compare directly with their approach.

\subsubsection*{Bridge-Finding and Biconnectivity}
The algorithmic study of biconnectivity dates back to Hopcroft and Tarjan~\cite{HT71,HT73} and Paton~\cite{Paton71}, who presented linear-time algorithms, based on depth-first search, for finding articulation points, bridges, and 2-connected components in undirected graphs. These graph-theoretic concepts are central in, e.g., planar graph recognition and drawing~\cite{Kant93}, or network analysis~\cite{Brandes04,Sariyuce13}.

Since depth-first search is inherently sequential~\cite{Reif85}, a parallel biconnectivity algorithm remained a challenge for a while. Following several less efficient parallel algorithms~\cite{Eckstein79,SJ81,TC84}, Tarjan and Vishkin~\cite{TV85} were the first to give an optimal, $O(\log n)$-time and $O(n + m)$-work, PRAM biconnectivity algorithm.

Follow-up works included heuristic improvements to the algorithm~\cite{CB05}, and evaluation on the \emph{Explicit Multi-Threading} (XMT) platform with promising results~\cite{EV12}. More recent papers~\cite{SM14, CK16} pointed out that Tarjan-Vishkin algorithm is not efficient on multi-core CPU. Instead, they suggested alternative heuristic solutions, without theoretical worst-case guarantees, based on breadth-first search and disregarding the Euler tour technique. Adhering to this trend, Wadwekar and Kothapalli~\cite{WK17} implemented a BFS-based biconnectivity algorithm on GPU, achieving significant speedups over multi-core CPU. That is the only GPU biconnectivity algorithm we are aware of.

\subsection{Experimental Setup}

We run our experiments on an NVIDIA GeForce GTX 980 GPU, with 2048 CUDA cores, and an Intel Xeon X5650 CPU, with 6 physical and 12 virtual cores.
The reported GPU performance ignores memory transfer times between CPU and GPU\footnote{Following the usual practice, we assume our algorithms are meant to be parts of a larger GPU processing pipeline, and thus the input data is already present in the GPU memory.}.
The source code, in C++ and CUDA,
is available at GitHub\footnote{\url{https://github.com/stobis/euler-meets-cuda/}}.

\section{The Euler Tour Technique}
\label{sec:EulerDescription}

The Euler tour technique is a classical PRAM method for representing and performing computations on trees. For problems where an input graph is not necessarily a tree (e.g.~the biconnectivity problem) the technique is usually applied to a spanning tree of the input graph.

The crux of the approach is to represent a (rooted) tree as a (singly linked) list of directed half-edges in the order of a depth-first search traversal (see Figure~\ref{fig:EulerTourExample}), i.e.~as an Euler tour.

Note that every subtree (a subgraph induced by a node and all of its descendants) corresponds to an interval in the list. Hence many node statistics can be easily calculated as prefix sums or range queries. For example, one can assign weight $1$ to each edge going down\footnote{Note that a directed half-edge goes down if and only if it appears before its twin (i.e.~the opposite direction half-edge) in the list.}, and weight $0$ to each edge going up, and compute the prefix sums to obtain the preorder numbering of the nodes. If the edges going up are instead assigned weights $-1$, the prefix sums are equal to the node levels (i.e.~distances from the root).

In the PRAM model such prefix sums can be computed with a simple modification of a list-ranking algorithm, which can be implemented in optimal $O(\log n)$ time on $O(n/\log n)$ processors~\cite{Cole86,Anderson88}. For a practical GPU algorithm, however, we need a more careful approach (see Section~\ref{sec:usingeuler}).

\begin{figure*}
    \begin{minipage}[t]{0.38\linewidth}
        \resizebox{\linewidth}{!}{\definecolor{c1}{RGB}{55,126,184}

\begin{tikzpicture}
  \begin{scope}[level distance=1.5cm,
          level 1/.style={sibling distance=2.2cm},
          level 2/.style={sibling distance=1.7cm}]
      \tikzstyle{every node}=[circle,draw]
    \node[inner sep=0.1em](1){\footnotesize 0}
      child {
              node [inner sep=0.1em] (2) {\footnotesize 2}
              child { node [inner sep=0.1em] (5) {\footnotesize 1} }
              child { node [inner sep=0.1em] (6) {\footnotesize 5} }
          }
      child {
              node [inner sep=0.1em] (3) {\footnotesize 3}
          }
      child {
              node [inner sep=0.1em] (4) {\footnotesize 4}
          };

      \draw[->, color=c1, very thick] ($(1)+(-.3,.1)$) -- node[above,midway,draw=none]{1} ($(2)+(-0,.3)$);
      \draw[->, color=c1, very thick] ($(2)+(-.3,0)$) -- node[above,midway,draw=none]{2} ($(5)+(-.15,.3)$);
      \draw[->, color=c1, very thick] ($(5)+(.3,0)$) -- node[below,midway,draw=none]{3} ($(2)+(-.05,-.6)$);
      \draw[->, color=c1, very thick] ($(2)+(.05,-.6)$) -- node[below,midway,draw=none]{4} ($(6)+(-0.3,0)$);
      \draw[->, color=c1, very thick] ($(6)+(.15,.3)$) -- node[right,midway,draw=none]{5} ($(2)+(.35,-.05)$);
      \draw[->, color=c1, very thick] ($(2)+(.45,0)$) -- node[below,midway,draw=none]{6} ($(1)+(-.3,-.55)$);
      \draw[->, color=c1, very thick] ($(1)+(-.25,-.6)$) -- node[below left,midway,draw=none]{7} ($(3)+(-.25,.2)$);
      \draw[->, color=c1, very thick] ($(3)+(.25,.2)$) -- node[below right,midway,draw=none]{8} ($(1)+(.25,-.6)$);
      \draw[->, color=c1, very thick] ($(1)+(.3,-.55)$) -- node[below,midway,draw=none]{9} ($(4)+(-.45,0)$);
      \draw[->, color=c1, very thick] ($(4)+(-0,.3)$) -- node[above,midway,draw=none]{10} ($(1)+(.3,.1)$);

      \draw[dashed, color=c1] ($(2)+(-0,.3)$) to[out=210, in=60] ($(2)+(-.3,0)$);
      \draw[dashed, color=c1] ($(5)+(-.15,.3)$) to[out=230, in=180] ($(5)+(-.1,-.22)$) to[out=-30, in=250] ($(5)+(.3,0)$);
      \draw[dashed, color=c1] ($(2)+(-.05,-.6)$) to[out=90, in=90] ($(2)+(.05,-.6)$);
      \draw[dashed, color=c1] ($(6)+(-.3,0)$) to[out=-70, in=-150] ($(6)+(.1,-.22)$) to[out=0, in=-50] ($(6)+(.15,.3)$);
      \draw[dashed, color=c1] ($(2)+(.35,-.05)$) to[out=0, in=180] ($(2)+(.45,0)$);
      \draw[dashed, color=c1] ($(1)+(-.3,-.55)$) to[out=0, in=90] ($(1)+(-.25,-.6)$);
      \draw[dashed, color=c1] ($(3)+(-.25,.2)$) to[out=-90, in=190] ($(3)+(0,-.25)$) to[out=-10, in=-90] ($(3)+(.25,.2)$);
      \draw[dashed, color=c1] ($(1)+(.3,-.55)$) to[out=180, in=90] ($(1)+(.25,-.6)$);
      \draw[dashed, color=c1] ($(4)+(-.45,0)$) to[out=-30, in=-135] ($(4)+(.2,-.16)$) to[out=50, in=-30] ($(4)+(0,.3)$);

  \end{scope}
\end{tikzpicture}}
        \caption{Example of an Euler tour}
        \label{fig:EulerTourExample}
    \end{minipage}
    \hfill
    \begin{minipage}[t]{0.58\linewidth}
        \resizebox{\linewidth}{!}{
\definecolor{twinEdgeColor}{RGB}{0,128,0}
\definecolor{nextEdgeColor}{RGB}{255,0,0}

\begin{tikzpicture}[
labelnode/.style={rectangle, draw=white!0, fill=white, thick, minimum size=1mm},
squarednode/.style={rectangle, draw=black!20, fill=white, thick, minimum size=5mm},
squarednodefirst/.style={rectangle, draw=orange!60, fill=yellow!5, thick, minimum size=5mm},
lolz/.style={edge, draw=orange!60, fill=yellow!5, thick, minimum size=5mm},
align=center,node distance=1cm and 0.0cm
]
\node[labelnode] (labelA) at (-1,0) {$A$:};
\node[squarednode]          (edgeA1)                             {(0,2)};
\node[squarednode]          (edgeA2)        [right=of edgeA1]    {(2,0)};
\node[squarednode]          (edgeA3)        [right=of edgeA2]    {(0,3)};
\node[squarednode]          (edgeA4)        [right=of edgeA3]    {(3,0)};
\node[squarednode]          (edgeA5)        [right=of edgeA4]    {(0,4)};
\node[squarednode]          (edgeA6)        [right=of edgeA5]    {(4,0)};
\node[squarednode]          (edgeA7)        [right=of edgeA6]    {(2,1)};
\node[squarednode]          (edgeA8)        [right=of edgeA7]    {(1,2)};
\node[squarednode]          (edgeA9)        [right=of edgeA8]    {(2,5)};
\node[squarednode]          (edgeA10)       [right=of edgeA9]    {(5,2)};

\node[labelnode] at (-1,0) [below=of labelA] {$B$:};
\node[squarednodefirst]     (edgeB1)        [below=of edgeA1]    {(0,2)};
\node[squarednode]          (edgeB2)        [right=of edgeB1]    {(0,3)};
\node[squarednode]          (edgeB3)        [right=of edgeB2]    {(0,4)};
\node[squarednodefirst]     (edgeB4)        [right=of edgeB3]    {(1,2)};
\node[squarednodefirst]     (edgeB5)        [right=of edgeB4]    {(2,0)};
\node[squarednode]          (edgeB6)        [right=of edgeB5]    {(2,1)};
\node[squarednode]          (edgeB7)        [right=of edgeB6]    {(2,5)};
\node[squarednodefirst]     (edgeB8)        [right=of edgeB7]    {(3,0)};
\node[squarednodefirst]     (edgeB9)        [right=of edgeB8]    {(4,0)};
\node[squarednodefirst]     (edgeB10)       [right=of edgeB9]    {(5,2)};

\draw[->, twinEdgeColor, semithick] (edgeA1.north)    to [out=30,in=150]      (edgeA2.north);
\draw[->, twinEdgeColor, semithick] (edgeA2.south)    to [out=-150,in=-30]    (edgeA1.south);

\draw[->, twinEdgeColor, semithick] (edgeA3.north)    to [out=30,in=150]      (edgeA4.north);
\draw[->, twinEdgeColor, semithick] (edgeA4.south)    to [out=-150,in=-30]    (edgeA3.south);

\draw[->, twinEdgeColor, semithick] (edgeA5.north)    to [out=30,in=150]      (edgeA6.north);
\draw[->, twinEdgeColor, semithick] (edgeA6.south)    to [out=-150,in=-30]    (edgeA5.south);

\draw[->, twinEdgeColor, semithick] (edgeA7.north)    to [out=30,in=150]      (edgeA8.north);
\draw[->, twinEdgeColor, semithick] (edgeA8.south)    to [out=-150,in=-30]    (edgeA7.south);

\draw[->, twinEdgeColor, semithick] (edgeA9.north)    to [out=30,in=150]      (edgeA10.north);
\draw[->, twinEdgeColor, semithick] (edgeA10.south)   to [out=-150,in=-30]    (edgeA9.south);

\draw[->, nextEdgeColor, semithick] (edgeB1.north)    to [out=30,in=150]      (edgeB2.north);
\draw[->, nextEdgeColor, semithick] (edgeB2.north)    to [out=30,in=150]      (edgeB3.north);
\draw[->, nextEdgeColor, semithick] (edgeB3.south)    to [out=-150,in=-30]    (edgeB1.south);
\draw[nextEdgeColor, semithick]     (edgeB4)          edge[loop above] node{} (edgeB4);
\draw[->, nextEdgeColor, semithick] (edgeB5.north)    to [out=30,in=150]      (edgeB6.north);
\draw[->, nextEdgeColor, semithick] (edgeB6.north)    to [out=30,in=150]      (edgeB7.north);
\draw[->, nextEdgeColor, semithick] (edgeB7.south)    to [out=-150,in=-30]    (edgeB5.south);
\draw[nextEdgeColor, semithick]     (edgeB8)          edge[loop above] node{} (edgeB8);
\draw[nextEdgeColor, semithick]     (edgeB9)          edge[loop above] node{} (edgeB9);
\draw[nextEdgeColor, semithick]     (edgeB10)         edge[loop above] node{} (edgeB10);

\matrix [below left,ampersand replacement=\&,nodes in empty cells,nodes={minimum height=#1,
anchor=center}] (mat) at (current bounding box.south east) {
 \node [name=mat-1-1, minimum size=4mm] {}; \& \node [name=mat-1-2, label=right:twin] {}; \&
 \node [minimum size=5mm] {}; \&
 \node [name=mat-1-3, minimum size=4mm] {}; \& \node [name=mat-1-4, label=right:next] {}; \&
 \node [minimum size=5mm] {}; \&
 \node [squarednodefirst] {}; \& \node [label=right:first] {}; \\
};
\draw[->, twinEdgeColor, thick] (mat-1-1.west) to [out=30,in=150] (mat-1-1.east);
\draw[->, nextEdgeColor, thick] (mat-1-3.west) to [out=30,in=150] (mat-1-3.east);

\end{tikzpicture}
}
        \caption{Intermediate DCEL-like representation of the tree from Figure~\ref{fig:EulerTourExample}}
        \label{fig:EulerTourExampleArrays}
    \end{minipage}
\end{figure*}

\subsection{Constructing an Euler Tour}
Since a tree is rarely already present in the Euler tour representation, an important part of the technique is constructing a tour in the first place. 

The actual construction algorithm has to depend on the input tree representation. Here we present an algorithm working with a very unstructured input: an unordered collection of undirected edges, represented as pairs of node identifiers. Other common tree representations can be easily and quickly converted to this representation. The first step in the Euler tour construction is creating an intermediate DCEL-like\footnote{DCEL stands for \emph{doubly connected edge list}, which is a common data structure for planar graphs and geometric problems.} representation of the tree. We remark that, unless the tree is already in such DCEL-like representation, it is not clear if any additional structure in the input can significantly speed up the conversion, in particular by avoiding the costly sorting.

\subsubsection*{DCEL} Our intermediate goal is to represent the input $n$-node tree as a collection of $2\cdot(n-1)$ directed (half-)edges, with each edge storing two pointers: \emph{next} and \emph{twin}. For each vertex $x$, all edges $(x,y)$ starting at $x$ are organized in a singly-linked cyclic list, so that the \emph{next} pointer of an edge points to the next edge on that list. The \emph{twin} pointer of each edge $(x,y)$ points to the reverse direction edge $(y,x)$.

\subsubsection*{Constructing DCEL}
First, we create an array of directed half-edges $A$: For each undirected edge $\{x, y\}$ we put $(x, y)$ and $(y, x)$ next to each other in $A$. Then, we create the lexicographically sorted copy of $A$, which we call $B$. When sorting, we ensure that each element of either array keeps an up-to-date pointer to its copy in the other array. To finish the construction we observe that the \emph{twin} pointer of an edge shall point to its neighbor in $A$, and the \emph{next} pointer -- to its neighbor in $B$, unless it is the last edge starting at a given vertex $x$. In order to handle such edges, we need one additional array $first$, with $first[x]$ pointing to the first edge starting at $x$ in $B$. Then, whenever $B[i]=(x,y)$ and $B[i+1]=(x',y')$ for $x \neq x'$, the \emph{next} pointer of $(x,y)$ shall point at $first[x]$. See Figure~\ref{fig:EulerTourExampleArrays}.

\subsubsection*{From DCEL to an Euler Tour} Constructing an Euler tour, as a linked list, is now straightforward: The successor of an edge $e$ in the list, $\mathrm{succ}(e)$, is given by
\[\mathrm{succ}(e) = \mathrm{next}(\mathrm{twin}(e)).\]
Conceptually, after traversing an edge $e=(x,y)$ and arriving at vertex $x$, one looks where they came from, i.e.~$\mathrm{twin}(e) = (y,x)$, and departs using the next edge, i.e.~$\mathrm{next}(\mathrm{twin}(e))$. For example, looking at Figure~\ref{fig:EulerTourExample}, $\mathrm{succ}(6) = \mathrm{next}(\mathrm{twin}(6)) = \mathrm{next}(1) = 7$.

Technically, a list created in such a way is cyclic. In order to perform, say, a prefix-sum computation we need to split the list at some point. We do it at an arbitrary edge $(r,y)$ leaving the root $r$ of the tree. To put it differently, if we start with an unrooted tree, we choose the root by choosing the list head.

\subsection{Using an Euler Tour on GPU}
\label{sec:usingeuler}

List-ranking is a common building block for PRAM algorithms (see~\cite{Jaja92} for an overview). In its basic variant, the list-ranking operation takes as input a list of elements, and computes for each element its \emph{rank}, i.e.~the distance from the head of the list. It can be implemented in optimal $O(\log n)$ time on $O(n/\log n)$ processors~\cite{Cole86,Anderson88}.
Moreover, a simple adaptation of the basic algorithm allows to compute the prefix sum of arbitrary values stored in the list, without increasing the asymptotic complexity. That happens to be the same complexity as that of computing the prefix sum in an array. That is, in the theoretical PRAM model, the prefix sums in an array and in a list cost the same.

In practice, however, on GPU the scan operation (i.e.~array prefix sum) is much faster then list-ranking (see e.g.~\cite{WJ10}, who report 7-8x difference). Since most Euler tour applications involve several prefix sum computations on the tour, we propose the following important optimization. After calculating an Euler tour as a list, we call a single list-ranking on the list, and we use its output to create an array of (directed) edges in the Euler tour ordering. That allows us to perform all the following prefix sum calculations on the Euler tour by using a fast scan primitive on the array. Moreover, knowing the indices of an edge and its twin, we can determine if the directed edge goes up or down. As a result we can, e.g., easily determine parents of all nodes, which we do in the \emph{hybrid} algorithm proposed at the end of Section~\ref{sec:bridges_exp}.

We use the \emph{moderngpu} library~\cite{Bax16} for fast \emph{sort} (in DCEL construction) and \emph{scan} primitives. Using the library throughout the implementation saves us the burden of low-level fine tuning, typically required to achieve a good performance on GPU. For list-ranking we implement the GPU-optimized Wei-JaJa algorithm~\cite{WJ10}, which performs much better than the classical pointer jumping technique.

\section{First Application: Lowest Common Ancestors}
\label{sec:lca}

For a rooted tree $T$ and two nodes $x, y \in T$, the lowest common ancestor of $x$ and $y$ is the unique node furthest from the root which is an ancestor of both $x$ and $y$. Equivalently, it is the only node on the unique simple path from $x$ to $y$ which is an ancestor of both $x$ and $y$.

Given a rooted tree $T$ and a sequence of queries $(x_1,y_1), (x_2,y_2), \ldots, (x_q,y_q) \in T \times T$, the LCA problem asks to compute for each query $(x_i,y_i)$, $i\in\{1,2,\ldots,q\}$, the lowest common ancestor of $x_i$ and $y_i$.

\subsection{Algorithms}

There is a diverse collection of algorithms computing LCA with a (near-)linear preprocessing time and a constant or logarithmic query time, see e.g.~\cite{Fischer2006}. Two main approaches emerge among them: (1) an approach based on a tree decomposition, e.g.~\cite{SV88}, and (2) an approach based on the reduction to the range minimum query (RMQ) problem, e.g.~\cite{BFC00}.

The Inlabel algorithm~\cite{SV88} falls into the first category. It has a linear preprocessing time and a constant query time. Moreover, it was designed to be easily parallelizable. With the Euler tour technique, the preprocessing can be implemented to run in $O(\log n)$ time on $O(n/\log n)$ processors in the PRAM model. Each query can be then answered in constant time on a single processor, i.e.~a collection of $q$ queries can be answered in $O(1)$ time on $O(q)$ processors. Besides the Euler tour technique the algorithm requires no design changes to implement it on GPU. 

In order to select a single-core CPU baseline, we run a preliminary experiment with a sequential implementation of the Inlabel algorithm against a simple RMQ-based algorithm -- a variant of~\cite{BFC00}, using a segment tree and without the preprocessed lookup tables for all short sequences. The RMQ-based algorithm has a faster preprocessing, by a factor of two, and the Inlabel algorithm answers queries faster, by a factor of three. When the number of queries equals the number of nodes, the two algorithms perform on par with each other. We chose the Inlabel algorithm as our single-core CPU baseline. We also implemented, using OpenMP, the parallel Inlabel algorithm as our multi-core CPU baseline (we are not aware of any publicly available multi-core LCA algorithm).

\subsubsection*{The Inlabel Algorithm}
In a full binary tree the lowest common ancestor of nodes $x$ and $y$ can be found by examining the longest common prefix of the inorder numbers of $x$ and $y$. That can be done with a constant number of bitwise operations. The Inlabel algorithm~\cite{SV88} generalizes that observation to general trees.

Given a rooted tree $T$, the algorithm assigns to each node an \emph{inlabel} number, so that the two conditions are satisfied:
\begin{itemize}
    \item \emph{Path partition property.} The inlabel numbers partition $T$ into paths going top-down, each path consisting of nodes with the same inlabel number.
    \item \emph{Inorder property.} Let $B$ be the smallest full binary tree having at least $|T|$ nodes. We identify each node of $B$ with its inorder number. The inlabel numbers map the nodes from the input tree $T$ to the nodes of $B$, not necessarily injectively, so that for every $x,y \in T$, if $x$ is a descendant of $y$ (in $T$) than $\mathrm{inlabel}(x)$ is a descendant of $\mathrm{inlabel}(y)$ (in $B$).
\end{itemize}

Schieber and Vishkin~\cite{SV88} showed how to calculate the inlabel numbers, from the preorder numbers and subtree sizes, in $O(1)$ time per node, in parallel. Next, they showed how -- having calculated the inlabel numbers and two other auxiliary arrays -- to answer LCA queries in constant time, on a single processor per query. We refer to the original paper~\cite{SV88} for a detailed description.

The Euler tour technique can be used to calculate the preorder numbers, subtree sizes, and node depths, in $O(\log n)$ time and $O(n)$ total work. The remaining part of the preprocessing runs in $O(1)$ time and $O(n)$ total work. Apart from the Euler tour technique itself, adapting the algorithm from PRAM to GPU is straightforward.

\subsubsection*{The Na\"ive Algorithm}
Martins et al.~\cite{MT12} propose a na\"ive LCA algorithm for GPU: Each thread gets assigned a single LCA query $(x,y)$, and it traverses the tree upwards from $x$ and $y$, node by node, until the two paths meet.

Each query takes time proportional to the distance from $x$ to $y$, compared to $O(1)$ time of the theoretically optimal Inlabel algorithm. However, if one expects that typical instances have short $x$-$y$ paths, e.g. the trees are shallow, then the na\"ive algorithm benefits from its extreme simplicity and low constant factors.

In principle it is possible to implement the algorithm without any preprocessing at all. However, that would then require threads to mark visited nodes and, as a consequence, to issue scattered writes to large linear-size arrays -- a thing to avoid for an efficient implementation. Fortunately, a simple and effective preprocessing lets us handle queries in a constant memory.

The preprocessing amounts to calculating the \emph{level} of each node, i.e.~the distance from the root. We adapt a standard pointer jumping technique for this task:
Each node gets an \emph{ancestor} pointer, initially pointing to its parent, and level $0$ is assigned to the root. Then, until all levels are calculated, the ancestor pointers are updated so that their lengths double, i.e.~the ancestor of a node is updated to be the ancestor of its ancestor. The level of a node is calculated once its ancestor pointer points at the root or other node with an already known level. The preprocessing runs in $O(\log n)$ time and $O(n \log n)$ total work.

We remark that this is not a theoretically optimal way to compute the node levels, but our experiments show that is already fast enough never to be a bottleneck. We do however employ a simple practical optimization: We perform five jumps for each pointer in parallel, before synchronizing the threads globally, as this empirically proves to be faster than synchronizing after each parallel pointer jump. 

To handle a query $(x, y)$, a thread starts with two pointers, pointing at $x$ and $y$ at the beginning. If the levels of $x$ and $y$ are different, the pointer with the higher level is moved upwards, node by node, the number of times equal to the difference of levels, so that both pointers point at nodes at the same level. Then, both pointers are move upwards simultaneously, until they point at the same node, which is $\mathrm{LCA}(x, y)$. 

\subsection{Datasets}
We generate synthetic datasets for our experiments. That gives us a flexibility to gradually change an isolated parameter and observe its impact on the algorithms' behavior. Besides, to the best of our knowledge, there is no established dataset for the LCA problem.

We start with a method to generate trees of a logarithmic depth. Let $[n]=\{1,2,\ldots,n\}$ be the node set. Node $1$ is the root, and the parent of node $i$, for each $i \geq 2$, is selected uniformly at random from $\{1,\ldots,i-1\}$. Trees generated that way have the expected average node depth equal to $\ln n$.

In order to be able to generate deeper trees, we introduce an additional parameter, \emph{grasp}, which we denote by $\gamma$. Now, the parent of node $i$, for each $i \geq 2$, is selected uniformly at random from $\{\max(i-\gamma, 1), \ldots, i-1\}$. For example, $\gamma=1$ deterministically yields a path, and $\gamma=\infty$ recovers the initial shallow tree distribution. In general,
\[\mathbb{E}(\text{average node depth}) = \begin{cases}
    \ln n, & \text{if } \gamma = \infty, \\ 
    \frac{1}{\gamma+1} n + O(1), & \text{otherwise.}
\end{cases}\]

The above model lets us control the tree size and depth, but not other parameters, e.g.~the degree distribution. For our final LCA experiment we generate scale-free Barab{\'a}si-Albert trees~\cite{Barabasi99,Mori05}. The parent of node $i$ is again selected from $\{1,\ldots,i-1\}$, but with probabilities proportional to the degrees, instead of the uniform distribution. Such trees have power law degree distributions, resembling real-world networks, and are very shallow on average.

At the end, we map all node identifiers through a random permutation of $[n]$, so that the tree structure is maintained but the identifiers do not leak any information. We sample queries uniformly at random from $[n]\times[n]$.

The input is given to the algorithms as an array of parents -- i.e.~node $P[i]$ is the parent of node $i$, for every $i$ except for the root -- and an array of queries. We remark that such tree description gives a head start to the na\"ive algorithm, while the Inlabel algorithm is essentially indifferent to the input format.

\newgeometry{left=1cm, right=1cm, top=2.5cm, bottom=2.5cm}
\begin{figure}
    \centering
    \subfloat[Preprocessing shallow trees]{
        \input{img/_file-E1_grasp--1x=Ny=NdivPreprocessinglines_xM.pgf}
        \label{fig:LCAPreUnlimited}
    }%
    \hfill
    \subfloat[Preprocessing deep trees]{
        \input{img/_file-E1_grasp-1000x=Ny=NdivPreprocessinglines_xM.pgf}
        \label{fig:LCAPre1000}
    }

    \subfloat[Answering queries in shallow trees]{
        \input{img/_file-E1_grasp--1x=Ny=numQdivQuerieslines_xM.pgf}
        \label{fig:LCAQueUnlimited}
    }
    \hfill
    \subfloat[Answering queries in deep trees]{
        \input{img/_file-E1_grasp-1000x=Ny=numQdivQuerieslines_xM.pgf}
        \label{fig:LCAQue1000}
    }
    \caption{General comparison of LCA algorithms}
    \label{fig:LCA_main}
    
    \vspace*{2\floatsep}%
    
    \begin{minipage}{0.48\linewidth}
    \input{img/_file-E4_grasp--1x=numQy=Overalllines_xM.pgf}
    \caption{Combined preprocessing and query time, depending on number of queries (8M nodes, shallow)}
    \label{fig:QtoN}
    \end{minipage}
    \hfill
    \begin{minipage}{0.48\linewidth}
    \input{img/_file-E3x=AvgHeighty=Overalllines.pgf}
    \caption{Time to answer 8M LCA queries in 8M-node tree, depending on tree depth (including preprocessing)}
    \label{fig:AvgHei}
    \end{minipage}
\end{figure}%
\restoregeometry

\subsection{Experiments and Results}
For each set of parameters (i.e.~the number of nodes, number of queries, and grasp) that we considered, we generated five instances with different random seeds. Then, we run each algorithm on each instance five times. That gives $25$ runs for each data point, and we report the average running times over those runs. The standard deviation, over the $25$ runs, never exceeded $6\%$ for the GPU algorithms, $13\%$ for the single-core CPU algorithm and $18\%$ for the multi-core CPU algorithm. The standard deviation between the five runs on the same instance was below $6\%$ for all the algorithms.

\subsubsection*{General Comparison} In our first experiment we generated trees of size varying from one million to $32$ million nodes, and fed them to all four algorithms: (1)~the baseline single-core CPU implementation of the Inlabel algorithm, (2) the multi-core CPU Inlabel algorithm, (3) the supposedly practical na\"ive GPU algorithm, and (4) the theoretically optimal GPU implementation of the Inlabel algorithm. For each tree size we considered shallow trees ($\gamma=\infty$) and deep trees ($\gamma=1000$). We set the number of queries to be equal to the number of nodes. The throughput of the algorithms, separately for the preprocessing and queries, is presented in Figure~\ref{fig:LCA_main}.

Let us first look at the shallow trees (Figures~\ref{fig:LCAPreUnlimited} and~\ref{fig:LCAQueUnlimited}). Both GPU algorithms offer a speedup over the CPU baselines. Unsurprisingly, the na\"ive algorithm has the fastest preprocessing, and the parallel Inlabel algorithm answers queries fastest. Which of the two algorithms is faster in total will depend on the ratio of the tree size to the number of queries (see Figure~\ref{fig:QtoN}, which we analyse later).

Now, let us move our attention to the deep trees (Figures~\ref{fig:LCAPre1000} and~\ref{fig:LCAQue1000}). The preprocessing performance does not differ significantly compared to the shallow trees. However, the query performance of the na\"ive GPU algorithm degrades heavily, to the point that it even becomes slower than the single-core CPU baseline. The other three implementations, all based on the Inlabel algorithm, behave similarly as for the shallow trees.

Regardless of the tree depth, the GPU implementation of the Inlabel algorithm consistently outperforms its reference single-threaded CPU implementation, offering a speedup in the range of 4-11x for preprocessing and 22-52x for queries. The multi-threaded CPU implementation situates between the other two.

Two parameters -- the number of queries relative to the number of nodes, and the tree depth -- determine whether the na\"ive algorithm outperforms the GPU Inlabel implementation. In our next two experiments we explore the parameter regimes in which each of the two algorithms thrives.

\subsubsection*{Queries-to-Nodes Ratio}
In our next experiment we focused on shallow trees (which give the most advantage to the na\"ive algorithm), and varied the queries-to-nodes ratio. Specifically, we fixed the number of nodes to $8$ million, and varied the number of queries from one million to $128$ million (thus varying the queries-to-nodes ratio from $0.125$ to $16$). We report the total running times, including preprocessing and answering all queries, in Figure~\ref{fig:QtoN}. The parallel Inlabel algorithm begins to outperform the na\"ive algorithm at around 4-to-1 queries-to-nodes ratio. Since the throughput of both algorithms does not depend much on the number of nodes, i.e. the lines plotted in Figures~\ref{fig:LCAPreUnlimited} and~\ref{fig:LCAQueUnlimited} are roughly horizontal, we expect that the intersection point would fall in similar places also for other tree sizes.

\subsubsection*{Tree Depth}
Next, we explored how the tree depth impacts the algorithms' performance. We fixed the number of nodes and the number of queries, both to 8 million, and varied the grasp parameter from $1$ to $10^7$, effectively generating trees with the average node depth ranging from $16$ to $4\cdot10^6$. We report the total running times, including preprocessing and answering all queries, in Figure~\ref{fig:AvgHei}.

As we already could expect from the initial general comparison experiment, the running time of the GPU Inlabel implementation was consistent across tree depths, and equaled to about 350 milliseconds. The na\"ive algorithm was 2.6x faster on the least deep trees. However, already at $91$ average node depth the two algorithms drew, and the na\"ive algorithm's performance degraded very quickly with the increasing depth.

\subsubsection*{Batch Size}
\begin{figure}
    \centering
    \input{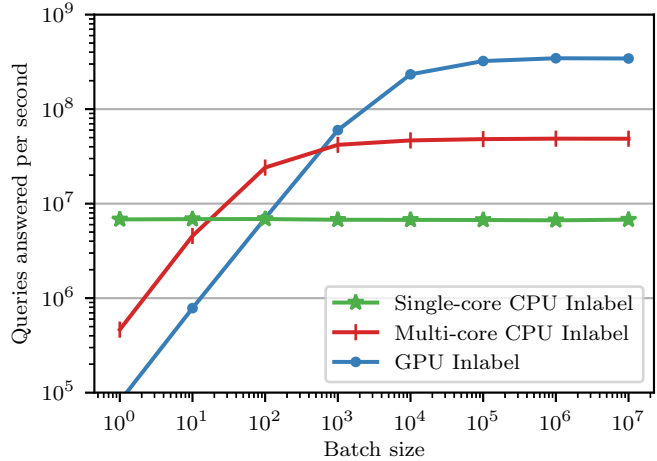}
    \caption{Benefit of answering LCA queries in parallel depending on batch size}
    \label{fig:LCA_batch}
\end{figure}
The LCA algorithms which we consider in this paper can work \emph{online}, i.e.~they can preprocess a tree without knowing the queries in advance, and then they can efficiently answer queries one by one. That has a benefit of, e.g., a small latency between providing a query and getting the answer for that query, and the ability to adapt future queries based on answers to previous queries. On the other hand, in order to benefit from solving the LCA problem on a parallel machine, such as multi-core CPU or GPU, the machine has to work on many queries at once.

In our next LCA experiment we explore the scenario in which not all queries are known beforehand yet they can be grouped into batches of certain size, and each batch can be given to an LCA algorithm at once. We are interested in how the batch size impacts the performance of the Inlabel algorithm.

We fixed the number of nodes to 8 million, and grasp to infinity. We generated 10 million random queries, split them into batches, and fed them to each algorithm batch by batch. The batch size varied from $1$ to $10^7$. Unsurprisingly, as batch size grew, the query time on multi-core CPU and GPU decreased (see Figure~\ref{fig:LCA_batch}). GPU was faster than single-core CPU with as few as 100 queries in a batch, and achieved its maximal throughput already at the batch size equal to 10\,000. Multi-core CPU beat single-core already just after 10 queries in a batch, and plateaued at around 1000 queries in a batch, where GPU took the lead.

\begin{figure}[t]
    \begin{adjustwidth}{-1.2cm}{-0.8cm}
    \centering
    \begin{minipage}{0.48\linewidth}
    \input{img/_file-E5x=Ny=NdivPreprocessinglines_xM.pgf}
    \caption{Preprocessing scale-free trees}
    \label{fig:LCAPreSF}
    \end{minipage}
    \hfill
    \begin{minipage}{0.48\linewidth}
    \input{img/_file-E5x=Ny=numQdivQuerieslines_xM.pgf}
    \caption{Answering queries in scale-free trees}
    \label{fig:LCAQueSF}
    \end{minipage}
    \end{adjustwidth}
\end{figure}%

\subsubsection*{Scale-Free Trees}
In our last LCA experiment we again test all four algorithms, this time on scale-free Barab{\'a}si-Albert trees. Similarly to our first LCA experiment (Figure~\ref{fig:LCA_main}), the number of nodes varies from one million to 32 million, the number of queries equals to the number of nodes, and we present the throughput of the algorithms separately for the preprocessing and queries, in Figures~\ref{fig:LCAPreSF} and~\ref{fig:LCAQueSF} respectively.

Unsurprisingly, these figures are very similar to Figures~\ref{fig:LCAPreUnlimited} and~\ref{fig:LCAQueUnlimited} for shallow trees, generated with $\gamma=\infty$. The only observable difference is the na\"ive GPU algorithm answering queries slightly faster, due to an even lower average node depth of the scale-free trees. The conclusions from our first LCA experiment remain valid also for the scale-free trees. This experiment confirms that the performance of the tested algorithms depends almost entirely on the tree size -- and does not depend on other parameters, such as the tree shape, degree distribution, etc. -- with the exception of the na\"ive GPU algorithm, whose query answering performance depends heavily also on the tree depth.

\section{Second Application: Bridges}
\label{sec:bridges}
In a connected undirected graph, a \emph{bridge} is an edge whose deletion makes the graph no longer connected. A \emph{2-edge-connected component} is a maximal subgraph which does not contain a bridge. A simple method to decompose a graph into 2-edge-connected components is to find all bridges, remove them, and find connected components in the resulting graph. Closely related notions of an \emph{articulation point} and a \emph{2-vertex-connected component} are defined similarly for vertices. 

For the sake of simplicity, in this paper we focus on the following problem: Given a graph $G=(V,E)$, determine for each edge $e \in E$ whether $e$ is a bridge. This basic problem already captures most of the combinatorial structure related to biconnectivity. In particular, the classical bridge-finding algorithm~\cite{HT73} uses the \emph{low} function, which is a central tool also for 2-edge- and 2-vertex-connected components decomposition, and which is tightly related to depth-first search trees -- a major obstacle for an efficient parallelization~\cite{Reif85}.

\subsection{Algorithms}
\label{sec:bridge_alg}
First we recall the classical (sequential) bridge-finding algorithm~\cite{HT71,HT73,Paton71}. Let us fix a spanning tree $T$, and let us identify the nodes with their preorder numbers. For a node $v \in T$, we define $\mathrm{low}(v)$ to be the minimum (preorder number) of endpoints of non-tree edges whose other endpoints belong to the subtree rooted in $v$. The low function can be easily computed in linear time. If $T$ is a depth-first search tree, then a tree edge\footnote{A non-tree edge is never a bridge.} $\{u, \mathrm{parent}(u)\}$ is a bridge if and only if $\mathrm{low}(u) \geq u$. Said differently, if it is possible to escape the subtree of $u$, it is possible to do so by going to a vertex smaller than $u$ (in the preorder numbering), which is specifically a property of DFS trees.

\subsubsection*{Tarjan-Vishkin (TV) Algorithm}
Tarjan~\cite{Tar74} proposed how to modify the above algorithm to work with any spanning tree -- thus escaping the DFS parallelization obstacle. Tarjan and Vishkin~\cite{TV85} showed how to implement the modified algorithm in parallel on PRAM.

In addition to the $low$ function, defined as above, the algorithm also computes the $high$ function, which is defined similarly but with maximum in place of minimum. For any spanning tree $T$, a tree edge $\{u, \mathrm{parent}(u)\}$ is a bridge if and only if at least one of $\mathrm{low}(u)$ and $\mathrm{high}(u)$ points outside of the subtree of~$u$, i.e.~outside of the interval $\big[\mathrm{preorder}(u), \mathrm{preorder}(u)+\mathrm{size}(u)\big)$, where $\mathrm{size}(u)$ denotes the size of the subtree.

The algorithm has three phases: (1) constructing a spanning tree, (2) rooting the tree and calculating required node statistics, and (3) computing the \emph{low} and \emph{high} functions in order to find bridges.

We use a GPU-optimized connected components algorithm by Jaiganesh and Burtscher~\cite{JB18}, which constructs a spanning tree as a byproduct.

Using the Euler tour technique, we root the tree, calculate the preorder numbers and subtree sizes, and for each node its minimum and maximum non-tree neighbor. For the last two values we use a specialized scan implementation \textit{segreduce} from the \textit{moderngpu} library~\cite{Bax16}.

In order to compute the \emph{low} and \emph{high} functions we need to aggregate, over subtrees, the per-node minimum and maximum non-tree neighbors. That tasks boils down to solving the range minimum query (RMQ) problem, which we do using the \emph{segment tree} data structure.

\subsubsection*{Chaitanya-Kothapalli (CK) Algorithm}
The alternative CK bridge-finding method is a simple, worst-case quadratic work, heuristic algorithm. It involves multiple walks on the edges of the graph, and works particularly well for graphs with a small diameter. Implementations of the CK algorithm are state-of-the-art parallel biconnectivity algorithms for multi-core CPU~\cite{CK16} and GPU~\cite{WK17}.

The algorithm consists of two phases. First, it finds a (rooted) spanning tree of the input graph. In the second phase, for each non-tree edge in parallel, it starts from its endpoints and walks up the tree up to their lowest common ancestor, marking tree edges visited along the way. A tree edge is a bridge if and only if it never get marked. For a detailed description and a proof of correctness see~\cite{CK16,WK17}.

A spanning tree algorithm for the first phase can be chosen arbitrarily, but a parallel BFS is used in most implementations~\cite{CK16,WK17}. The choice of BFS guarantees that the spanning tree depth is at most a factor of two from the minimum, which allows to bound the work performed in the marking phase by $O(md)$, where $d$ denotes the diameter. Moreover, BFS is a very well studied graph primitive, with highly-optimized implementations available for many parallel environments.

Our multi-core CPU implementation of the CK algorithm is based on the publicly available implementation of the runner-up algorithm~\cite{SM14b} using OpenMP. Our GPU implementation uses our own implementation of BFS, based on~\cite{MGG12} and using \textit{moderngpu} primitives~\cite{Bax16}.

\begin{table}[ht]
  \centering
  \begin{tabular}{@{}lrrrr@{}} \toprule
    Graph                  &Nodes &Edges & {Bridges} & {Diameter} \\ \midrule
    kron\_g500-logn16      & 55K  & 4.9M & 12K       & 6       \\
    kron\_g500-logn17      & 107K & 10M  & 26K       & 6       \\
    kron\_g500-logn18      & 210K & 21M  & 54K       & 6       \\
    kron\_g500-logn19      & 409K & 43M  & 113K      & 7       \\
    kron\_g500-logn20      & 795K & 89M  & 233K      & 7       \\
    kron\_g500-logn21      & 1.5M & 182M & 477K      & 7       \\
    \\
    web-wikipedia2009      & 1.8M & 9.0M & 1.4M      & 323     \\
    cit-Patents            & 3.7M & 33M  & 1.3M      & 26      \\
    socfb-A-anon           & 3.0M & 47M  & 3.3M      & 12      \\
    soc-LiveJournal1       & 4.8M & 85M  & 2.2M      & 20      \\
    ca-hollywood-2009      & 1.0M & 112M & 23K       & 12      \\
    \\
    USA-road-d.E           & 3.5M & 8.7M & 2.2M      & 4K      \\
    USA-road-d.W           & 6.2M & 15M  & 3.8M      & 4K      \\
    great-britain-osm      & 7.7M & 16M  & 4.8M      & 9K      \\
    USA-road-d.CTR         & 14M  & 34M  & 8.5M      & 6K      \\
    USA-road-d.USA         & 23M  & 58M  & 14M       & 9K      \\
    \bottomrule
  \end{tabular}
  \caption{Statistics of largest connected components in graphs used in bridge-finding experiments}
  \label{tab:bridges}
\end{table}

\subsection{Datasets}
We evaluated the bridge-finding algorithms on 16 graphs, listed in Table~\ref{tab:bridges}. They can be split into three main categories: (1) synthetic Kronecker graphs~\cite{Le08}, which model moderately sparse networks with small diameters, (2) real-world Internet and social network graphs, with similar characteristics to Kronecker graphs, and (3) real-world road graphs, which are extremely sparse and have significantly larger diameters. The graphs are available for download from public repositories~\cite{DIMACS9,DIMACS10,SNAP,NR}, and they commonly appear in experimental papers on biconnected components, e.g.~\cite{CK16,WK17}, and GPU graph algorithms in general, e.g.~\cite{Gun17}. We preprocessed each graph to keep only its largest connected component. Table~\ref{tab:bridges} summarizes basic statistics of the final test instances.

\subsection{Experiments and Results}
\label{sec:bridges_exp}
\begin{wrapfigure}{r}{0.45\textwidth}
    \vspace{-.8cm}
    \centering
    \input{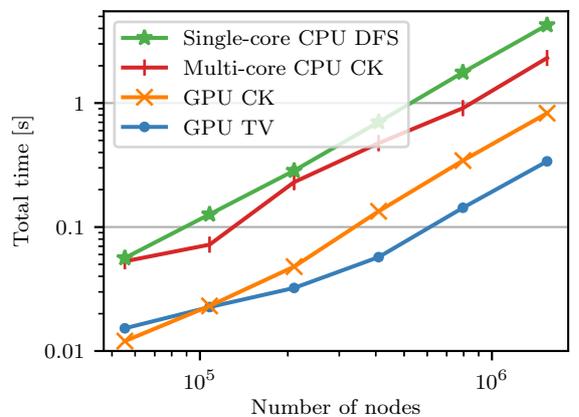}
    \caption{Comparison of bridge-finding algorithms on Kronecker graphs}
    \label{fig:bridgesKron}
    \vspace{-.5cm}
\end{wrapfigure}

We ran each algorithm on each graph 5 times, and we report the average running times. The standard deviation was below 5\% for the four algorithms discussed in Section~\ref{sec:bridge_alg}, i.e.~TV and CK algorithms on GPU, and DFS and CK on CPU. For the hybrid algorithm, proposed at the end of this section, the standard deviation was below 8.5\%.

\begin{figure}[t]
    \centering
    \input{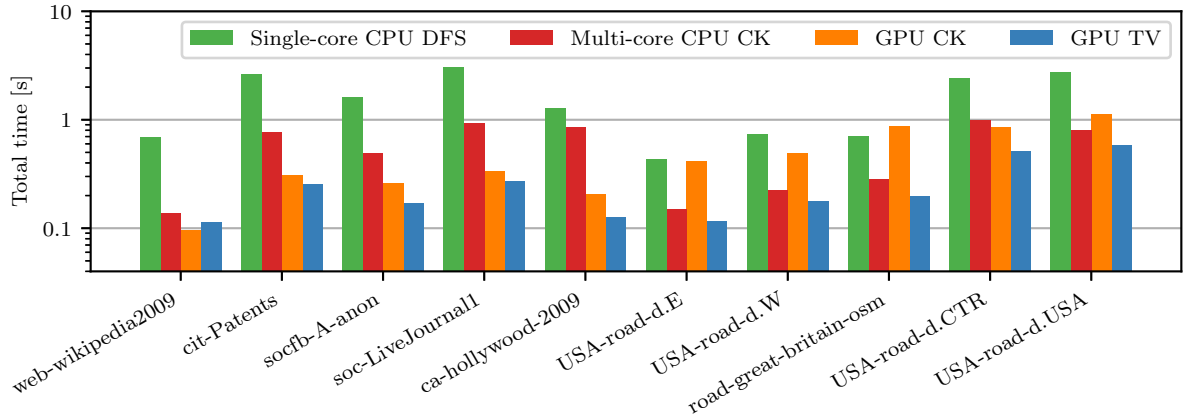}
    \caption{Comparison of bridge-finding algorithms on real-world graphs}
    \label{fig:bridgesOverall}
\end{figure}

First, let us just focus on the total running times, to see which of the algorithms performs best. See
Figure~\ref{fig:bridgesKron} for the performance on Kronecker graphs, and Figure~\ref{fig:bridgesOverall} for real-world graphs.

Comparing the two GPU algorithms, we see that TV is faster than CK on all instances but two: the smallest Kronecker graph, and the Wikipedia graph. The difference in favor of TV is most significant for the road graphs with large diameter, where TV is up to 4.7x faster than CK.

When compared to the CPU baselines, the TV algorithm shows 4-12x speedups over the single-core DFS implementation and 1.25-8x speedups over the multi-core CK implementation. A speedup of less than 3x over the multi-core algorithm can be observed only for the road graphs and the small Wikipedia graph.

\begin{figure}
    \centering
    \input{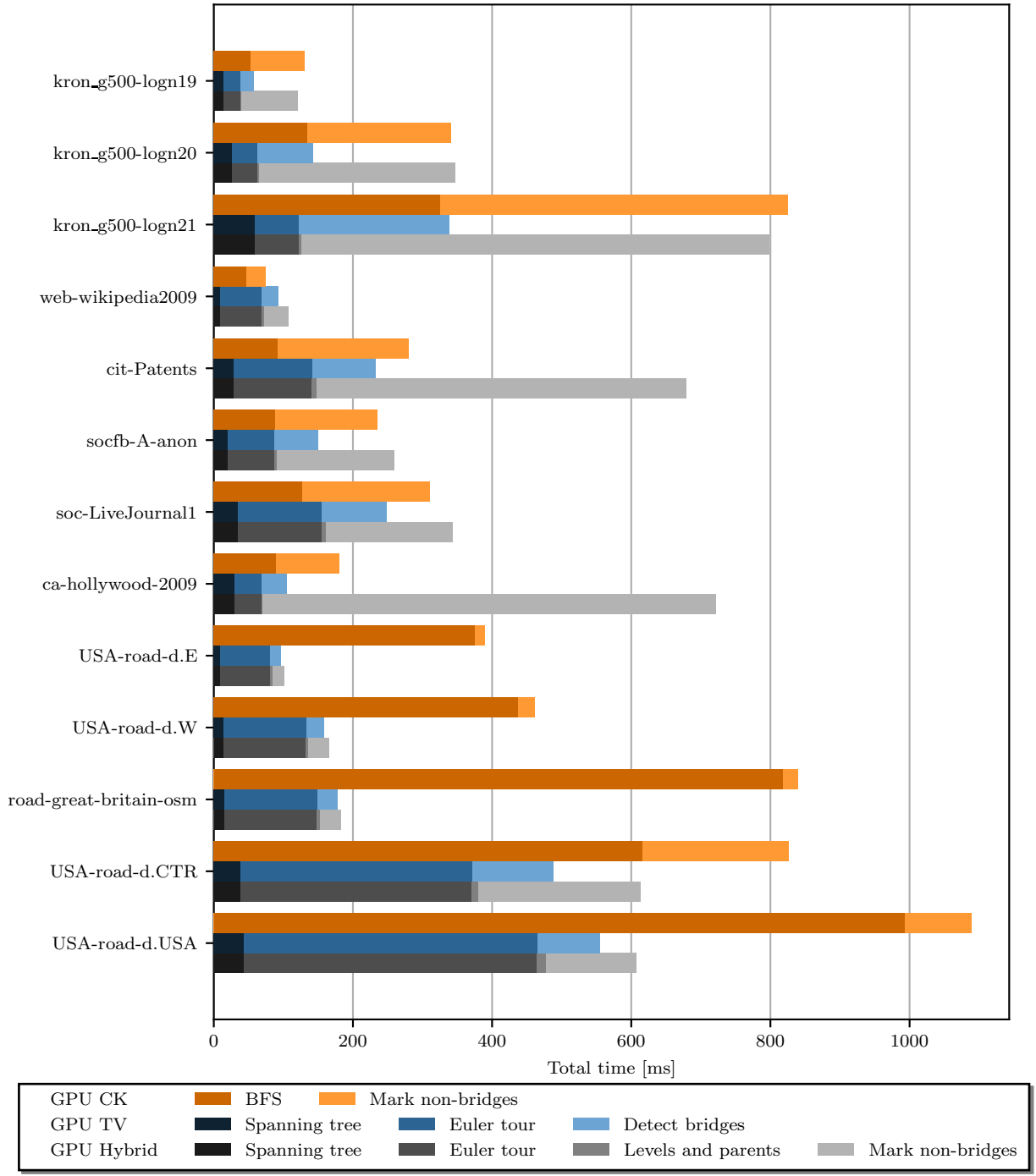}
  \caption{Running time breakdown of GPU bridge-finding algorithms}
  \label{fig:bridgesDetailed}
\end{figure}

\subsubsection*{The Hybrid Algorithm}
In order to better understand the performance of the algorithms we can look at Figure \ref{fig:bridgesDetailed}, which presents what contributes to their running times.  For the GPU CK implementation BFS always accounts for a significant portion of the running time, and becomes an apparent bottleneck when the input diameter grows. That is not surprising given the nature of parallel BFS performance, which is very sensitive to the diameter (see e.g.~\cite{MGG12}).

The correctness of the marking phase of the CK algorithm does not depend on specific properties of breadth-first search trees. Hence, a natural attempt to speed up the algorithm is to replace BFS with a different, faster algorithm computing a spanning tree. Such tree is likely to be deeper than a BFS tree, but the hope is that the performance of the marking phase degrades only slightly, and the time saved on finding a spanning tree is not entirely lost.

We propose to compute spanning trees with the same connectivity algorithm~\cite{JB18} which we use in the TV implementation. It is important to note that this algorithm outputs an unrooted spanning tree, but the marking phase requires a rooted tree, i.e. each node has to know its parent. Additionally, each node also needs to know its level (distance from the root). We compute both parents and levels using the Euler tour technique.

Looking again at Figure \ref{fig:bridgesDetailed}, we can see that the hybrid algorithm is unlikely to outperform TV, for the following reason. Both algorithms begin with spanning tree and Euler tour computations. Then, the hybrid algorithm still has to execute the marking phase, which is likely to be no faster than the marking phase of CK, which in turn is (on most of the test instances) slower than the last remaining stage of TV.

We evaluated the hybrid algorithm experimentally -- as predicted, it
was often faster then CK, but it never outperformed TV. We leave it for further research whether there is a faster method to get a rooted spanning tree, without resorting to the Euler tour technique, and whether a resulting hybrid algorithm could beat TV.

\section{Conclusions}
We propose how to adapt to GPU a classical building block of optimal parallel graph algorithms, the Euler tour technique. As a proof of concept we study two natural graph problems for which, to the best of our knowledge, the only previous GPU algorithms are based on simple heuristics, efficient for typical problem instances.
In our experiments the Euler tour-based algorithms outperform the previous approaches unless the tree is shallow and the queries are few, in case of the LCA problem, and unless the graph is small, in case of bridge-finding.

\section*{Acknowledgments}
We thank anonymous reviewers for many helpful suggestions and comments.
We also thank Grzegorz Gutowski for his indispensable support in running the experiments.

\bibliographystyle{plainurl}
\bibliography{IEEEabrv,main}

\clearpage

\end{document}